\documentclass[12pt]{article}
\usepackage{graphicx}
\usepackage{color}

\usepackage{amsmath,amssymb}
\def\sl2R{sl(2,\mathbb{R})}
\def\SL2R{SL(2,\mathbb{R})}

\def\11{\mbox{$1$}}

\newcommand{\rref}[1]{(\ref{#1})}
\newcommand{\beqn}{\begin{equation}}
\newcommand{\eeqn}{\end{equation}}
\newcommand{\beqarr}{\begin{eqnarray}}
\newcommand{\eeqarr}{\end{eqnarray}}

\newcommand{\matc}{\begin{array}{c}}
\newcommand{\matcc}{\begin{array}{cc}}
\newcommand{\matccc}{\begin{array}{ccc}}
\newcommand{\matcccc}{\begin{array}{cccc}}
\newcommand{\emat}{\end{array}}

\newcommand{\IH}{\relax{\rm I\kern-.18em H}}
\newcommand{\IR}{\relax{\rm I\kern-.18em R}}
\newcommand{\IK}{\relax{\rm I\kern-.18em K}}
\newcommand{\II}{\hbox{\rm 1\kern-.28em I}}
\newcommand{\Is}{\relax{\rm 1\kern-.35em 1}}
\begin{document}

\begin{titlepage}
\pagestyle{empty}
August 2004
~\hfill RU-04-4-B~~~~
\vskip .9in

\begin{center}
\textbf{\Large Strings, Dipoles and Fuzzy Spheres}
\vskip .6in
{\large Bogdan Morariu\footnote{morariu@summit.rockefeller.edu}
\vskip .4in
{\em Department of Physics,        Rockefeller University   \\
 New York, NY 10021, USA }}
\end{center}
\vskip .5in

\abstract{I discuss a scaling limit, where open strings in the WZW-model
behave as dipoles with charges confined to a spherical brane and 
projected to the lowest Landau level. Then I show how the joining 
and splitting interactions of these dipoles are naturally described 
using the fuzzy sphere algebra.}

\end{titlepage}

\newpage
\renewcommand{\thepage}{\arabic{page}}

\setcounter{page}{1}
\setcounter{footnote}{0}

\section{Introduction}
\label{Intro}
The idea that space coordinates might be noncommutative is due to
Heisenberg. Later Peierls showed that the coordinates of particles
in strong magnetic fields become noncommutative upon projection 
to the lowest Landau level (LLL). However, this realization of
noncommutativity is phenomenological and led Snyder to
propose a {\em fundamental} quantization of space coordinates.  
In more recent times this proposal was successfully implemented for
gauge theories on noncommutative spaces such as the noncommutative
plane, torus~\cite{Conn} or sphere~\cite{Madore:1991bw}. 

This begs the question as to why such
deformations of ordinary quantum field gauge theory exist and whether
this is related to the original phenomenological realization of
noncommutativity as a projection to the LLL in strong magnetic fields. 
Indeed, in the flat space case the low energy effective description of
strings in a strong NS-NS background is given by noncommutative gauge
theory~\cite{Connes:1997cr,Schomerus:1999ug,Chu:1999qz,
Schwarz:1998qj,Brace:1999ku}. For reviews see~\cite{Seiberg:1999vs,DN,Sza}.
It was emphasized in~\cite{Bigatti:1999iz} 
that at low energy
strings behave as dipoles with the charges in the LLL and the joining
and separation interaction of these dipoles is naturally described by
gauge theory on the noncommutative plane~\cite{Yin:1999ba,Bigatti:1999iz}. 
For further detailed studies see 
also~\cite{Kiem:2001pw,Kiem:2001fn,Kiem:2001dk,Kiem:2001du}.

In this paper I explore a
similar limit for open strings on group manifolds. I will concentrate
on the ${\rm SU}(2)$ group but most of the analysis can be generalized to an
arbitrary group $G$\,. Note that $AdS_3\times S^3 \times T^4$ is an
exact string theory background, thus the ${\rm SU}(2)$ WZW-model 
corresponds to the $S^3$ factor and can be embedded in string theory.
Then, the D2-branes on which strings end are
ordinary two spheres. I show that there exists a scaling limit similar to
the one in flat space and then explore how the fuzzy sphere algebra arises
naturally. The resulting noncommutative gauge theory on a fuzzy sphere
was first derived by Alekseev, Recknagel and Schomerus 
in~\cite{Alekseev:1999bs}.
However, the derivation uses conformal field theory results and does
not reveal the simple geometric picture of 
interacting dipoles which immediately leads to the fuzzy sphere algebra.
Our derivation identifies the states
of the string at low energy with the wave function of the dipole in
the LLL. Furthermore, we also show that a remarkable formula used to describe
multiplication on the fuzzy sphere has the direct physical
interpretation as an interaction vertex.

The paper is organized as follows. 
Section~\ref{Review} contains a review of D-branes and open strings in the
WZW-model.
In section~\ref{Scaling}, I introduce
a scaling limit and obtain the low energy action. I quantize the
reduced action in section~\ref{Fuzzy} and show how the joining and
splitting interaction is naturally described by the fuzzy sphere algebra.
Finally, in the appendix I show how Kaluza-Klein reduction can be used 
to obtain the quantum states of the particle in a monopole magnetic field. 

\section{D-branes in the WZW-model}
\label{Review}
\setcounter{equation}{0}
In this section we review D2-branes and open strings in the  $SU(2)$ 
level $k$ WZW-model. This subject has a long history starting with  
Ishibashi and Cardy~\cite{Ishibashi:1988kg,Cardy:ir} who studied
consistent boundary states in the WZW-model using current algebra techniques. 
In~\cite{Klimcik:1996hp} an action approach was used to show that
D2-branes with magnetic flux and lying on conjugacy classes give
a consistent extension of the WZW-model to open strings. It was
realized in \cite{Alekseev:1998mc} that the Cardy states are nothing
but the D2-brane conjugacy classes introduced in~~\cite{Klimcik:1996hp}.
The relation of these D-branes to fuzzy spheres was first noted
in~\cite{Alekseev:1999bs} and the noncommutative gauge field theory
describing the low energy excitations of open strings was written down
in~\cite{Alekseev:2000fd}.
Finally  in~\cite{Bachas:2000ik}, using  the Born-Infeld action it
was shown that these spherical D2-branes are stable.
In this section I will review the WZW action of open strings on group
manifolds following~\cite{Klimcik:1996hp}\,, but using a different
notation in order to agree with the
standard conventions. 

\subsection{Closed strings}
Before I discuss D-branes and open strings let me first consider
closed strings.
Just as the action of a particle coupled to a gauge field $A_{\mu}$ is not 
gauge invariant, the string coupled to the $B_{\mu\nu}$
field does not have a gauge invariant action. To obtain a gauge
invariant action one has to consider a closed path for the particle 
or a closed world sheet for the string. In
the Minkowski formulation one can take the difference between two
world sheets with the same initial and final string configurations. 
Let $g : \Sigma \rightarrow G$ denote an arbitrary map from the 
world sheet to the target group manifold $G$ connecting the initial and
final string configurations. Also choose a fixed reference map $g : \Sigma_0
\rightarrow G$ connecting the same initial and final string
configurations\footnote{The use of the same letter $g$ is
intentional, as $g$ denotes a single function defined on $\Sigma -
\Sigma_0$}.  We can now form a gauge invariant combination from the WZW 
actions ${\cal S}_{\Sigma}$ and ${\cal S}_{\Sigma_0}$ given by 
\begin{equation}
{\cal S}_{\Sigma} - {\cal S}_{\Sigma_0} =
\frac{k}{8 \pi}
\oint_{\Sigma - \Sigma_0} \! \! d^2\sigma \, {\rm Tr}(g^{-1}\partial^{\mu} g
g^{-1}\partial_{\mu} g)+
\frac{k}{12 \pi}
\int_{\cal B} {\rm Tr}\left[ (\tilde{g}^{-1}d\tilde{g})^{\wedge
3}\right]~.
\label{WZW}
\end{equation}
Here ${\cal B}$ is any $3$-dimensional manifold such that its
boundary $\partial{\cal B}$ satisfies
$\partial {\cal B}=\Sigma -\Sigma_0$\,, and $\tilde{g}$ is an arbitrary map
$\tilde{g} : {\cal B} \rightarrow G$ such that it coincides with $g$ when
restricted to the boundary $\Sigma -\Sigma_0$\,. 
Note that since the second homology group $H_2(G)$ is trivial it is 
always possible to construct the map $\tilde{g}$\,.
The second term is the
nonlinear sigma model and the last term is the Wess-Zumino action. The
only ambiguity in~\rref{WZW} is in the choice of the manifold ${\cal
B}$ and of the map~$\tilde{g}$\,. The difference 
between two such choices  in \rref{WZW} is the integral of 
the $3$-form $H = 
\frac{k}{12\pi} {\rm Tr}\left[ (\tilde{g}^{-1}d\tilde{g})^{\wedge
3}\right]$ on a closed $3$-dimensional sub-manifold of $G$. If the
level $k$ is an integer the ambiguity in the action is a multiple of
$2\pi$ which is allowed quantum mechanically~\cite{Witten:1983ar}. 
Mathematically, this is
the statement that  
$H \in {\rm H}^3(G, \mathbb{Z})$\, i.e. 
$H$ is an integral cohomology cycle.

\subsection{Open strings}
In the open string case the endpoints of the string are constrained to
live on D-branes which are sub-manifolds of $G$\,. We further assume
that there exist $U(1)$ gauge fields localized on the D-branes. In
this case $\Sigma -\Sigma_0$ has two closed boundaries $\Delta_1$ and
$\Delta_2$ localized on the
D-branes. If we further assume that
the D-branes are simply connected $\Delta_1$ and
$\Delta_2$ are contractable to a point. Let $\Sigma_1$ and $\Sigma_2$
denote 2-manifolds with boundaries  $\Delta_1$ and
$\Delta_2$\,.   Then $\Sigma -\Sigma_0 + \Sigma_1 -\Sigma_2$ is a
closed 2-dimensional surface, 
as shown in the figure below.
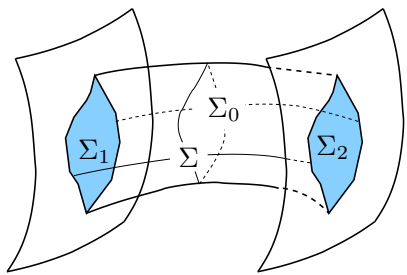
\vspace{0.3cm}
\begin{center}
\includegraphics[scale=0.35]{D2brane.0}
\end{center}
\vspace{0.3cm}
Let ${\cal B}$\, denote any 
3-dimensional manifold such that 
$\partial{\cal B}=\Sigma -\Sigma_0 + \Sigma_1 -\Sigma_2$\,
and let $B$ denote a two form such that 
$dB =H$\,. If $F$ is the field strength of the $U(1)$ gauge field
in the D-brane we  can define the open version of~\rref{WZW}
\begin{eqnarray}
{\cal S}_{\Sigma} - {\cal S}_{\Sigma_0} & =&
\frac{k}{8 \pi}
\int_{\Sigma - \Sigma_0}
 \! \! d^2\sigma \, {\rm Tr}(g^{-1}\partial^{\mu} g
g^{-1}\partial_{\mu} g)+\nonumber \\
&~&
\frac{k}{12 \pi}
\int_{\cal B} {\rm Tr}\left[ (\tilde{g}^{-1}d\tilde{g})^{\wedge
3}\right]  \label{OWZW}+ \\
&~&   \int_{\Sigma_1} (F - B)- \int_{\Sigma_2} (F - B)~~.
\nonumber
\end{eqnarray}
Note that \rref{OWZW} is manifestly gauge invariant
since the gauge transformation $B' = B + d\Lambda$ is accompanied by
$A' = A+ \Lambda$, thus $F-B$ is gauge invariant. There are 
further ambiguities in \rref{OWZW} besides the ones found
in~\rref{WZW} related to
the choice of $\Sigma_1$ and $\Sigma_2$. Again quantum mechanics allows
for a multiple of $2\pi$ ambiguity in the action and this implies that
the integral on $F$ on any
closed two manifold must be an integer. 
In conclusion, we have found that the following topological quantizations
\begin{equation}
H \in {\rm H}^3(G, \mathbb{Z})~,~~F \in {\rm H}^2(G, \mathbb{Z})
\nonumber
\end{equation}
are the necessary and sufficient conditions for a consistent path
integral formulation of the open WZW-model.

\subsection{Born-Infeld stability analysis}
If D2-branes exist in the ${\rm SU}(2)$ level $k$ WZW-model their ground 
state must be $S^2$ by symmetry. But there are no nontrivial cycles 
with this topology in $G={\rm SU}(2)$ therefore there must 
exist a dynamical mechanism that stabilizes the branes.
Indeed Bachas, Douglas and Schweigert used 
the Born-Infeld effective action to show that the $S^2$ branes are 
stable~\cite{Bachas:2000ik}.
Let us briefly review their analysis. 

The metric in the $SU(2)$ level $k$ WZW-model
is the standard $S_3$ metric
\begin{equation}
ds^2 =
k \alpha' \,[ d\psi^2 + 
\sin^2 \! \psi \,(d\phi^2 + \sin^2 \! \theta \,d\phi^2)]~,
\nonumber
\end{equation}
and the NS-NS field strength is up to normalization the volume $3$-form on
$S_3$ given by 
\begin{equation}
H = dB =
\frac{k}{\pi} \,\sin^2\! \psi \,\sin \theta  \,\, d\psi\, d\theta\, d\phi ~.
\nonumber
\end{equation}
Consider a spherical D2-brane located at some fixed $\psi$
and carrying magnetic flux $m \in \mathbb{Z}$ satisfying $0< m< k$. The uniform gauge field is
\begin{equation}
F = dA = 
-\frac{m}{2}  \,\sin \theta  \, d\theta\, d\phi ~.
\label{Fbrane}
\end{equation}
The energy of such a configuration (to lowest order in $\alpha'$) 
obtained using the Born-Infeld action is given by 
\begin{eqnarray}
E_m(\psi) &=&
T_{(2)} 
\int_0^\pi \! d\theta \int_0^{2\pi}\! d\phi\,
\sqrt{\det(\hat{G}+2\pi\alpha'(\hat{B}+ F))}
\nonumber\\
&=& 4\pi k\alpha' \, T_{(2)} \left(
\sin^4\!\psi +(\psi - \frac{\sin 2\psi}{2}-\frac{\pi m}{k})^2
\right)^{1/2}~,\nonumber
\end{eqnarray}
where $\hat{G}$ and $\hat{B}$ are the induced metric and NS-NS 2-form
and $T_{(2)}$ is the D2-brane tension.
There is a unique minimum of the energy for $0< m < k$ given by 
\begin{equation}
\psi_m = \frac{\pi m}{k}~.
\nonumber
\end{equation}

Furthermore, in~\cite{Bachas:2000ik} it was also shown 
that these spherically symmetric 
configurations are stable against small fluctuations. A quadratic
expansion around the spherical D-brane contains only positive mass
terms except for three zero modes corresponding to translations of
locations of the center of mass away from pole at $\psi = 0$\,.

\section{Scaling limit}
\label{Scaling}
In this section I consider the scaling limit obtained by
taking $k$ to infinity and holding~$m$ fixed. First I will write down
the action for the $SU(2)$ level $k$ WZW-model describing
open strings ending on the D2-brane located at $\psi_m$ in a fixed gauge
\begin{eqnarray}
{\cal S} &=&  \! \int  d \tau \int_{-l/2}^{l/2} \! d\sigma\, \left.
~\frac{k}{2}\,  \right\{   [\dot{\psi}^2 +
\sin^2 \! \psi \,(\dot{\phi}^2 + \sin^2 \! \theta \,\dot{\phi}^2)]-
\label{kin} \\
&~&\,
\, [{\psi'}^{\,2} +
\sin^2 \! \psi \,({\phi'}^{\,2} + \sin^2 \! \theta \,{\phi'}^{\,2})]+
\label{pot}\\
&~& \, \left.
 \, \left(\psi -\frac{\sin(2\psi)}{2} \right)\,
 \sin\theta \,
(\dot{\theta}\,\phi' -\theta'\,\dot{\phi})  \right\} + 
\label{Bfield}\\
&~& \int  d \tau\, \frac{m}{2}\, ( \cos\theta_+ \,\dot{\phi_+}
-\cos\theta_- \,\dot{\phi_-} )~. \label{Aterm}
\end{eqnarray}
The terms~\rref{kin} and~\rref{pot} are the nonlinear $\sigma$-model
part of the action. The term~\rref{Bfield} is a local parameterization
of the WZ-action with the B-field nonsingular at $\psi=0$\,. The final
term~\rref{Aterm} gives the coupling of the two ends of the string to
the gauge field on the D2-brane~\rref{Fbrane}.

First do the
following rescaling
\begin{equation}
\left\{
\begin{array}{ccc}
\psi &=& \frac{2\pi}{k}\,r~, \\
\tau &=& \frac{k}{2\pi} t~.
\end{array}
\right. \label{Scale}
\end{equation}
To motivate this rescaling first note that large $k$ is the
semi-classical limit and we expect string excitations to decouple.
Since the brane is located at $\psi_m = \frac{\pi m}{k}$ and the
string $\psi$ coordinate must be of order $\psi_m$ it is useful to
introduce a rescaled coordinate $r$ which will be finite. 
The stable brane labeled by $m$ is
now located at 
\begin{equation}
r_m = \frac{m}{2}~. \nonumber
\end{equation}
The rescaled time $t$ is introduced
so that the low energy states of the string are also of order
one. Upon inserting~\rref{Scale} into the WZW action, the
terms~\rref{kin} and \rref{Bfield} scale as $k^{-2}$ while the 
terms~\rref{pot} and \rref{Aterm} scale as  $k^{0}$\,. Thus as $k
\rightarrow \infty$ the WZW action reduces to 
\begin{eqnarray}
{\cal S} & = &
\int \! d t\,  \frac{m}{2}\, ( \cos(\theta_+) \dot{\phi_+} 
-\cos(\theta_-) \dot{\phi_-} )- \label{DEP} \\
&~&\int \!  d t
\int_{-l/2}^{l/2} \! \! d\sigma\,
 \frac{1}{2}\, [{r'}^{\,2} +
r^2 \,({\theta'}^{\,2} + \sin^2 \! \theta \,{\phi'}^{\,2})]~.\label{HAM}
\end{eqnarray}
Note that the quadratic kinetic term~\rref{kin} has disappeared and 
the action is already in Hamiltonian form. However now only the
endpoints of the string have conjugate canonical momenta. 
The bulk of the string coordinates are auxiliary fields 
and can be integrated out.
For large $k$ the $S_3$ sphere becomes very large and the
strings see a flat metric. That is why I used $r$ to denote
the rescaled coordinate $\psi$\,.  In Cartesian coordinates, the
equations of motion derived from~\rref{HAM} are just $x_a''=0$\,, thus
the string is just a straight line segment connecting two points on the 
$D2$-brane. Integrating these equations and plugging the solution back
in~\rref{HAM} we obtain the Hamiltonian
\begin{equation}
{\cal H} = \frac{1}{2 l} \Delta^2~, \label{Ham}
\end{equation}
where $\Delta$ denotes the length of the string. 

To review, I have found that the large $k$ limit of the 
WZW open strings with the rescaling~\rref{Scale} 
behave like dipoles whose charges are
connected by an elastic string of string constant
$l^{-1}$. Furthermore, 
no mass term is present for the charges therefore upon
quantization only the LLL will be present.

\section{The fuzzy sphere algebra}
\label{Fuzzy}
In this section I will present a brief review of the fuzzy sphere
algebra. Consider
the following angular momentum truncation of functions of the $S^2$\,
sphere
\begin{equation}
f(\theta, \phi) =\sum_{l=0}^{m} \sum_{p=-l}^{l} C_p^l Y_l^p(\theta, \phi)~.
\label{trunc} 
\end{equation}  
The space of truncated functions is denoted $S^2_{m+1}$ and
could be used to obtain a finite number of degrees of
freedom for a field theory on $S^2$. However, the set of
truncated functions do not form an algebra with respect to the usual function
multiplication so it appears difficult to write any interaction
terms without higher angular momenta resurfacing. 

Fortunately a modified multiplication exists which makes the set of truncated
functions~\rref{trunc} into an algebra. It is defined as
follows. First recall that the spherical harmonics  
are traceless homogeneous polynomials in the normalized 
Cartesian coordinates $x_a /r$. Thus one can rewrite~\rref{trunc} as
\begin{equation}
f=c^0 +c_a^1 \frac{x^a}{r} + c^2_{ab} \frac{x^a x^b -\delta^{ab}
r^2}{r^2} 
+ \ldots~, \label{symm}
\end{equation}
where the coefficients $C^l_m$ in~\rref{trunc} and the symmetric
traceless coefficients $c^l_{a_1\ldots a_l}$
in~\rref{symm} are linearly related.
Let $J^a$ denote a $N=m+1$ dimensional representation of the $su(2)$ Lie
algebra and define $X^a = f J^a$ where $f^2 = 4 r^2/(N^2 -1)$. Then
$X^a$ satisfy
\begin{eqnarray}
[X^a, X^b] &=& i f \epsilon^{abc} X^c~,~~
X^2 = r^2~. \nonumber
\end{eqnarray}

We can now define a linear map ${\cal M}: S^2_N \longrightarrow {\rm Mat}(N)$ 
from the space of truncated
functions~\rref{trunc} to the space of $N$-dimensional matrices
\begin{equation}
x^{a_1} x^{a_2} \ldots x^{a_k} \stackrel{\cal M}{\longrightarrow} 
{\rm sym}(X^{a_1} X^{a_2} \ldots X^{a_k}) \nonumber
\end{equation}
which takes a product of $x^a$'s into the symmetrized product 
obtained by substituting  each $x^a$ by $X^a$.
The (complex) linear dimension of $S^2_N$\,
is just $\sum_{l=0}^{N-1} (2l+1)=N^2$\,, the same as the dimension 
of $N$-dimensional matrices. In fact 
the map ${\cal M}$ is one to one and onto. Therefore the product of
$n$-dimensional matrices
induces a ``star'' product on the space of truncated functions
\begin{equation}
f_1 * f_2 = {\cal M}^{-1}({\cal M}(f_1){\cal M}(f_1) )~.
\label{star}
\end{equation} 
The $*$-multiplication~\rref{star} makes  $S^2_N$\, into an algebra
called the fuzzy sphere algebra. 

For practical purposes it is better
to use the isomorphism ${\cal M}$ and think of $S^2_N$ as the set 
of N-dimensional matrices with the multiplication given by standard
matrix multiplication.

\section{Interacting dipoles and the fuzzy sphere}
\label{DIPOLES}
McGreevy, Susskind and Toumbas have performed a classical analysis of
the dipole model~\rref{DEP} in~\cite{McGreevy:2000cw} 
and conjectured that
the quantum version of the model must be described by noncommutative
geometry. Briefly, their analysis is as follows. As one increases the
angular momentum of the dipole the distance between the charges also
increases. Therefore the angular momentum must be cut off when the
size of the dipole equals the diameter of the sphere. It is this
angular momentum cutoff which is very suggestive of the fuzzy sphere
algebra. 
In the first part of the paper I have shown that the dipole
model~\cite{McGreevy:2000cw} gives the low energy description 
of the open WZW strings. In the remainder of the paper I will derive 
the fuzzy sphere
algebra by studying the interaction vertex of the quantized dipoles
thus confirming the McGreevy, Susskind and Toumbas conjecture.

Our first task is to quantize the dipole model~\rref{DEP}. It is convenient
to add a mass term to the two dipole charges, quantize the model and
then take the mass to zero. Then all the higher Landau levels
decouple. We will treat the Hamiltonian~\rref{Ham} as a perturbation
which will remove the degeneracy of the LLL.
Before discussing the dipole 
I will consider a single charged particle moving on $S^2$ in a
magnetic flux $m$\,.

Note that $S^2 = {\rm SU}(2)/{\rm U}(1)$ so we can identify functions on $S^2$
with ${\rm U}(1)$ left-invariant functions on ${\rm SU}(2)$\,. 
Let ${\cal D}_{pq}^{\,j}(g)$
denote the Wigner symbols defined as
\begin{equation}
{\cal D}_{pq}^{\,j}(g) = <p\,|U(g)|\,q>~, \label{WIGNER}
\end{equation}
where $|\,q>,~q=-j,\ldots,j$ are states in the spin $j$ representation
of $SU(2)$\,. For any $h={\rm diag}(e^{i\phi} , e^{-i\phi}) \in {\rm U}(1)$
we have
\begin{equation}
{\cal D}_{pq}^{\,j}(gh) = {\cal D}_{pq}^{\,j}(g) ~e^{i \,2q \,\phi}~.
\label{MOM}
\end{equation}
Thus, up to normalization we can identify the spherical harmonics
$Y_l^p$ with ${\cal D}_{p0}^{\,l}$ and expand functions on $S^2$ as
\begin{equation}
\psi(g) = \sum_{l=0}^{\infty}\sum_{p=-l}^{l} C_p^{l} {\cal
D}_{p0}^{\,l}(g)~.
\nonumber
\end{equation}

The other Wigner symbols ${\cal D}_{pq}^{\,j}$ for $q\neq 0$ are also 
useful~\cite{Haldane:1983xm}.
It turns out that the functions
\begin{equation}
\psi_{m}(g) = \sum_{j=q}^{\infty}\sum_{p=-j}^{j} C_p^{j} {\cal
D}_{pq}^{\,j}(g)~,
\label{MONO}
\end{equation}
where $m=2q \in \mathbb{Z}$\,, 
give the global expansion for sections of the charge $m$ monopole
line bundle.
From a mathematical point of view~\rref{MONO} is just a useful way of
representing section of a nontrivial line bundle as function on a
principal bundle. However, it is possible to arrive at this result, 
as I will briefly discuss by considering the auxiliary
problem of a free particle on $S^3$ followed by a Kaluza-Klein reduction.  
For the detailed calculations see the appendix.
 
To understand the above statement, consider a free particle
moving on $S^3 = {\rm  SU}(2)$. The isometry group is ${\rm SO}(4)$ which 
up to a global identification is the same as
${\rm SU}(2) \times \widetilde{{\rm SU}(2)}$\,. A complete set of commuting
generators is given by $J_3$\,,  $\widetilde{J}_3$ and $J^2 =
\widetilde{J}^2$\,.  Then $ {\cal D}_{mn}^{\,j}(g)$ form a complete
set of states with eigenvalues $m$\,, $n$\,, and $j(j+1)$\,.
Locally we can perform a Kaluza-Klein reduction along the ${\rm U}(1)$
factor. As shown in the appendix, 
the background metric $g_{AB}$ on $S^3$ decomposes as follows
\begin{equation}
(g_{AB})_{S^3}  ~ \longleftrightarrow ~(G_{\mu\nu}, A_{\mu},
\Phi)_{S^2}~.
\nonumber 
\end{equation}
Here $G_{\mu\nu}$ is the  metric on a $S^2$ sphere of diameter one, 
$A_{\mu}$ is the charge one 
monopole gauge field and $\Phi$ gives the size of the ${\rm U}(1)$ factor.

As usual in Kaluza-Klein reduction, fields carrying momentum in the
compact direction are charged under the  ${\rm U}(1)$ gauge fields
with charge proportional to the momentum $\widetilde{J}_3$\,. 
Therefore, using~\rref{MOM}
we see that~\rref{MONO} can be interpreted as the state of a particle 
of electric charge $m$
in a charge one monopole background. Equivalently we can
interpret it as the state of a charge one 
particle in a $m$ magnetic flux background.  The Hamiltonian for
the free particle on $S^3$ is proportional to $J^2 = j(j+1)$ so to
restrict~\rref{MONO} to the LLL we must set
$j=q$\,. 
The reduced Hilbert space is just
${\cal H}_m = \mathbb{C}^{m+1}$
and contains  only the states
\begin{equation}
\psi_m(g) = \sum_{p=-q}^{q} C_p {\cal
D}_{pq}^{\,q}(g)~.
\nonumber
\end{equation}

For a dipole the Hilbert space is given by
${\cal H}={\cal H}_m \times {\cal H}_{-m}=  \mathbb{C}^{N}\times 
\mathbb{C}^{N}$\, and this is isomorphic to the space
of $N$-dimensional matrices. This is the first indication that we are
on the right track given the relation between the fuzzy sphere algebra
and the matrix algebra.\footnote{If 
the string stretches between two D-branes with magnetic fluxes
$m$ and $m'$ centered at the same point, the Hilbert space is given by
${\cal H}_{mm'}={\cal H}_m \times {\cal H}_{-m'}=  \mathbb{C}^{N}\times 
\mathbb{C}^{N'}$\, and it is isomorphic to the space of $N\times N'$\,
dimensional matrices. These are the projective modules of the fuzzy
sphere algebra.} 
What remains to be shown is that matrix
multiplication is relevant for describing dipole
interactions. 
First
note that states of the dipole are of the form
\begin{equation}
\Psi(g_1,g_2) = \sum_{s,t= -q}^{q} M_{st} \,
{\cal D}_{-s -q}^{\,q}(g_1)\,{\cal D}_{t q}^{\,q}(g_2)\nonumber
\end{equation}
or using the unitarity of the representation 
\begin{equation}
\Psi(g_1,g_2) = \sum_{s,t= -q}^{q} M_{st} \,
\overline{{\cal D}_{s q}^{\,q}(g_1)}\,{\cal D}_{t q}^{\,q}(g_2)~.
\label{DIPO}
\end{equation}
It is useful to define an inner product
\begin{equation}
<\Psi^{(1)} \,|\, \Psi^{(2)}>=
(2q+1)^{2}\int\!\!\int_{\!{\rm SU}(2)}\!\!dg_1\,dg_2\, 
\overline{\Psi^{(1)}(g_1,g_2)}\,\Psi^{(2)}(g_1,g_2)~,
\label{INN}
\end{equation}
where $dg_i$ is the Haar measure on the ${\rm SU}(2)$ group.
Using the orthogonality relations for the Wigner symbols 
\begin{equation}
\int_{{\rm SU}(2)}\!\!dg\,
\overline{{\cal D}_{p_1q_1}^{\,j_1}(g)}\,{\cal D}_{p_2 q_2}^{\,j_2}(g)
=
\frac{1}{2 j_1 +1} \,\delta^{j_1 j_2}\, \delta_{p_1 p_2} \delta_{q_1 q_2}~,
\label{ORTHO}
\end{equation}
we see that $<\Psi^{(1)}\,|\,\Psi^{(2)}>={\rm Tr}(M^{(1)\dagger} M^{(2)})$\,.

Having performed first quantization for a single dipole we can now
consider the Fock space of multi-dipole states and introduce interactions.
Since the dipoles are the low energy
states of strings it is natural to assume a contact interaction for 
the endpoints of the dipoles i.e. the
positive charge of the first dipole annihilates the negative charge of
the second dipole etc. as shown below
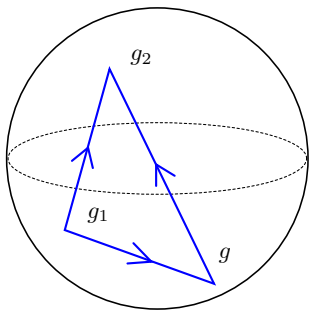
\vspace{0.3cm}
\begin{center}
\includegraphics[scale=0.25]{interaction.0}
\end{center}
\vspace{0.3cm}
For example, the vertex interaction for two
incoming and one outgoing dipoles is proportional to
\begin{equation}
<\Psi^{(3)}\,|\,\Psi^{(1)} * \Psi^{(2)}>~~
=~~~~~~~~~~~~~~~~~~~~~~~~~~~~~~~~~~~
~~~~~~~~~~~~~~~~~~~~~~~~~~~~~~~~~~~\nonumber
\end{equation}
\vspace{-0.5cm}
\begin{equation}
(2q+1)^{3}\int\!\!\int\!\!\int_{\!{\rm SU}(2)} \!dg_1\, dg_2\, dg_3\,
\overline{\Psi^{(3)}(g_1,g_3)}
\Psi^{(1)}(g_1,g_2)\Psi^{(2)}(g_2,g_3)~.\nonumber
\end{equation}
The charges of the dipoles have contact interactions and we
have to integrate over the locations of the interaction points. Note also
that the vertex interaction can be naturally written in terms of the
$*$-product 
\begin{equation}
\Psi^{(1)} * \Psi^{(2)}(g_1,g_2)=
(2q+1)\int_{\!{\cal SU}(2)}\!\!dg\, \Psi^{(1)}(g_1,g)\,\Psi^{(2)}(g,g_2)~,
\label{STAR}
\end{equation}
and the inner product~\rref{INN}\,.

Upon integration, using the orthogonality relations for the 
Wigner symbols~\rref{ORTHO}
the $*$-product~\rref{STAR}
reduces to
\begin{equation}
\Psi^{(1)} * \Psi^{(2)}(g_1,g_2)=
\sum_{s,t= -q}^{q} \,\left(
\sum_{u= -q}^{q} 
M^{(1)}_{su}M^{(2)}_{ut} \right)\,
\overline{{\cal D}_{s q}^{\,q}(g_1)}\,{\cal D}_{t q}^{\,q}(g_2)~.\nonumber
\end{equation}
Lo and behold, the state of the resulting dipole is described by the
product $M^{(1)}M^{(2)}$ 
of the matrices corresponding to the incoming dipoles.  
Thus I have shown that the vertex interaction
of dipoles is naturally described using traces and matrix
multiplication, the latter being equivalent to the $*$-multiplication 
of the fuzzy sphere algebra~\rref{star}. This is the main
result of the paper. 

Note that when the location of the two opposite charges of the dipole 
coincide the wave function is just $\Psi(g,g)$\,. This is a true function on
$S^2$ and can be identified with the truncation~\rref{trunc}\,.
There have been many attempts to write a star product on the fuzzy
sphere using~\rref{trunc}\,. 
In most cases however the resulting formulae lack the simplicity of the
Moyal star product on the plane. Alternatively one could use
a function of {\em two} variables $\Psi(g_1,g_2)$ as in~\rref{DIPO} in which
case the star product has the simple form~\rref{STAR}\,.
This remarkable form of representing the fuzzy sphere algebra was
introduced in~\cite{Nair:2001kr} as a trick in an attempt to write the
star product. We have found the physical interpretation of~\rref{DIPO}
as the state of the dipole with the charges on the LLL and of the 
star product~\rref{STAR}\, as the contact interaction of dipoles

Finally, note that the Hamiltonian~\rref{Ham} lifts the degeneracy of
the states~\rref{DIPO}\,. States with different total angular momentum have 
different energies. It is an interesting exercise to calculate 
these energies and the propagator.

\section{Concluding remarks}
\label{CR}
While our presentation highlighted the link between
the nonlocal nature of the dipole interaction and the fuzzy sphere
algebra I would like to point out that we have first taken the low energy
limit of the WZW-model and then quantized. In fact one should first quantize
and then take the low energy limit as quantization and taking the
scaling limit do not commute. Just as for the superstring in
flat space (after the GSO projection) the lowest energy states
correspond to some of the string oscillator state being excited, the
lowest energy states for the WZW open strings also have one oscillator
state excited. Since there are three possible oscillators there exist
three species of strings (similar to the polarizations states of open
strings in flat space). This is why in the matrix 
model~\cite{Alekseev:2000fd} one
needs three operators $X^i,~i=1, 2 ,3$\,. 

Our analysis also makes it clear why it is necessary to restrict to
D-branes of finite $m$ magnetic flux. If one takes $k \rightarrow
\infty$ keeping the ratio $m/k$ fixed it is not necessary to do the
rescaling of the coordinate $\psi$ as in~\rref{Scale} and then the
kinetic energy term does not vanish. In that case 
large angular momentum states have energies of the same order of
magnitude as stringy excitations. Thus the latter
do not decouple. This is the reason for the lack of associativity
observed in~\cite{Alekseev:1999bs,Steinacker:2001fj}.

A problem left for further study is to extend our approach
to D-branes on arbitrary group manifolds.\footnote{In general 
these are twined conjugacy classes.}
In particular it would be interesting to investigate nonabelian fibers
which are related to
the nonabelian quantum Hall effect~\cite{Zhang:2001xs}\,.

\section*{Acknowledgments}
I would like to thank Alexios Polychronakos for useful discussions.
This work was supported in part by the~ U.S.~ Department~ of~ Energy~
under ~Contract Number DE-FG02-91ER40651-TASK B.

\section*{Appendix}
In this appendix I will discuss the Kaluza-Klein reduction 
of the Schroedinger equation for a free particle from $S^3={\rm SU}(2)$ 
to $S^2= ${\rm SU}(2)/{\rm U}(1)\, along the ${\rm U}(1)$ fiber. 
States carrying momentum $m$ 
along ${\rm U}(1)$ will
be identified with a charged particle on $S^2$ in a magnetic flux $m$\,.

In a neighborhood on the identity
any $g\in {\rm SU}(2)$ can be written as 
$g(\theta,\phi,y)=g_0(\theta,\phi) h(y)$\,,
where $g_0(\theta,\phi)= e^{i\frac{\sigma_3}{2}\phi } 
e^{i\frac{\sigma_2}{2}\theta }$\, 
gives a local parameterization of the base and $h(y)=e^{i \sigma_3 y}$
a parameterization of the ${\rm U}(1)$ fiber. Plugging $g=g_0 h$ into 
the Killing metric on ${\rm SU}(2)$ given by $ds^2 = -\frac{1}{2}{\rm
Tr}(g^{-1}dg\,g^{-1}dg)$ we obtain
\begin{equation}
ds^2 = -\frac{1}{2}{\rm
Tr}[(g_0^{-1}dg_0 g_0^{-1}dg_0) +
2(g_0^{-1}dg_0 dh h^{-1})+
(h^{-1}dhh^{-1}dh)]~.\label{MET1}
\end{equation} 
On the other hand we can write the metric on $S^3$ in a  Kaluza-Klein form 
\begin{equation}
ds^2=
\left(
\begin{array}{ccc}
d\theta &
d\phi&
dy
\end{array}
\right)
\left( 
\begin{array}{cc}
1 & A \\
0 & 1
\end{array}
\right)
\left( 
\begin{array}{cc}
G & 0 \\
0 & \Phi
\end{array}
\right)
\left( 
\begin{array}{cc}
1 & 0 \\
A^T & 1
\end{array}
\right)
\left(
\begin{array}{c}
d\theta\\
d\phi\\
dy
\end{array}
\right)~,\label{MET2}
\end{equation}
where $G_{\mu\nu}$ is the metric on the $S^2$ base space, $A_{\mu}$ is
a ${\rm U}(1)$ gauge field and $\Phi$ is a scalar. 
We can obtain $G_{\mu\nu}$\,,  $A_{\mu}$ and  $\Phi$ by 
comparing~\rref{MET1} and~\rref{MET2}. After a trivial
calculation we have
\begin{equation}
 G_{\mu\nu} dx^{\mu}dx^{\nu} = \frac{1}{4}(d\theta^2 + \sin^2\theta
d\phi^2)
~,~~
A_{\mu}dx^{\mu} ~= \frac{1}{2}\cos\theta d\phi~,~~
\Phi~ = 1~,\nonumber
\end{equation}
that is, the metric on a $S^2$ of diameter one, the monopole gauge field and a
constant scalar field measuring the radius of the $U(1)$ fiber.
Note that $g_0$ is not well defined globally. However we can cover the base
with patches to obtain the global description.  
Locally different choices of $g_0$ are then $U(1)$ gauge equivalent.

Next consider a (non relativistic) scalar field on $S^3$  with the action
given by
\begin{equation}
{\cal S}=
\int dt\int d^3 x \sqrt{g} \,\left(\,
i \overline{\Psi}\partial_t \Psi - 
\frac{1}{2M}\, \partial_A  \overline{\Psi} g^{AB}(x) \partial_B
\Psi\,
- E_0 \overline{\Psi}\Psi\right)~, 
\label{SCH}
\end{equation}
where $g_{AB}$ is the standard $S^3$ metric. 
The equation of motion derived from~\rref{SCH} is just the Schroedinger
equation for a free particle of mass $M$\,,
and the last term is such that the 
ground state energy is set to $E_0$\,.

The isometry group is ${\rm SO}(4)$ which modulo some global
identifications
is just ${\rm SU}(2) \times \widetilde{{\rm SU}(2)}$\,.
A complete set of commuting
operators is given by $J_3$\,,  $\widetilde{J}_3$ and $J^2 =
\widetilde{J}^2$\,.  Then the Wigner symbols~\rref{WIGNER} 
form a complete set of states and satisfy
\begin{eqnarray}
 J_3 \,{\cal D}_{pq}^{\,j}(g) &=&  p\,{\cal D}_{pq}^{\,j}(g)~, 
\nonumber\\
 \widetilde{J}_3 \,{\cal D}_{pq}^{\,j}(g) &=&  q\,{\cal
D}_{pq}^{\,j}(g)~,
\nonumber\\
 J^2 \,{\cal D}_{pq}^{\,j}(g) &=&  j(j+1)\,{\cal D}_{pq}^{\,j}(g)~.
\nonumber
\end{eqnarray}
Furthermore, the Hamiltonian  which up to an additive constant is 
proportional to the Lapacian, can
be written as ${\cal H}_{S^3} = \frac{2}{M} J^a J^a +E_0=
 \frac{2}{M} j(j+1) +E_0$\,.

Locally we can perform the Kaluza-Klein reduction along the ${\rm U}(1)$
factor. If we substitute 
\begin{equation}
\Psi(\theta,\phi,y) =
\sum_{m\in\mathbb{Z}} \Psi_m(\theta,\phi) \frac{e^{imy}}{\sqrt{2\pi}}
\nonumber
\end{equation}
and the decomposition~\rref{MET2} of the metric into the
action~\rref{SCH}  we obtain
${\cal S} = \sum_{m\in\mathbb{Z}} {\cal S}_m$ where
\begin{eqnarray}
{\cal S}_m &=& 
\int \!dt \!\int \!d\theta d\phi \,\sqrt{G}  
 \left[ i\overline{\Psi}_m
\partial_t 
\Psi_m  \right.    \nonumber \\ 
&-& \left. \frac{1}{2 \cal M}~\overline{ D_{\mu}^{(m)}\Psi_m}G^{\mu\nu} 
D_{\mu}^{(m)}\Psi_m
-  \left( \frac{1}{2 M}  m^2+E_0\right) \, \overline{\Psi}_m
\Psi_m \right]~. \label{ACTION}
\end{eqnarray} 
Here $D_{\mu}^{(m)}=\partial_{\mu} - i m A_{\mu}$\, 
is the covariant derivative in
a charge $m$ monopole background. Thus fields carrying momentum
$m$ along the compact direction are charged under the ${\rm U}(1)$
gauge field with charge $m$\,.

Let us concentrate on the momentum
$m=2q$ sector for which a complete set of eigenfunctions is given by 
${\cal D}^{\,j}_{pq}(g)$.
Using~\rref{ACTION} we see that the Hamiltonian ${\cal
H}_{S^3}$ is related to the Hamiltonian 
${\cal H}_{S^2}^{(m)}$ of the particle on 
$S^2$ in a flux $m$ magnetic field by
${\cal H}_{S^3} ={\cal H}_{S^2}^{(m)} + \frac{1}{2 M} M^2+ E_0$\,. 
If we set $E_0 = \frac{m}{ M}$\, 
the eigenfunctions $D^{\,j}_{pq}(g)$ \,satisfy
\begin{equation}
{\cal H}_{S^2}^{(M)} {\cal D}^{\,j}_{pq}(g)= \frac{2}{M}[(j(j+1) -
q(q+1)] {\cal D}^{\,j}_{pq}(g)~.
\label{LL}
\end{equation}
Thus the lowest LLL $ {\cal D}_{pq}^{\,q}(g)~,~p=-q,\ldots,q$\, 
obtained for $j=q$\, have zero energy. As can be seen from~\rref{LL} 
in the limit $M \rightarrow \infty $\, all the higher Landau levels decouple.


\begin{thebibliography}{99}


\bibitem{Conn}
A.~Connes,
``Yang-Mills for noncommutative two-Tori,''
Operator Algebras and Mathematical Physics (Iowa City, Iowa, 1985)
237-266, Contemp. Math. Oper. Algebra. Math. Phys. 62, AMS 1987.





\bibitem{Madore:1991bw}
J.~Madore,
``The fuzzy sphere,''
Class.\ Quant.\ Grav.\  {\bf 9}, 69 (1992).


\bibitem{Connes:1997cr}
A.~Connes, M.~R.~Douglas and A.~Schwarz,
``Noncommutative geometry and matrix theory: Compactification on tori,''
JHEP {\bf 9802}, 003 (1998)
[arXiv:hep-th/9711162].


\bibitem{Schomerus:1999ug}
V.~Schomerus,
``D-branes and deformation quantization,''
JHEP {\bf 9906}, 030 (1999)
[hep-th/9903205].

\bibitem{Ardalan:1998ce}
F.~Ardalan, H.~Arfaei and M.~M.~Sheikh-Jabbari,
``Noncommutative geometry from strings and branes,''
JHEP {\bf 9902}, 016 (1999)
[arXiv:hep-th/9810072].

\bibitem{Chu:1999qz}
C.~Chu and P.~Ho,
``Noncommutative open string and D-brane,''
Nucl.\ Phys.\ B {\bf 550}, 151 (1999)
[hep-th/9812219];
``Constrained quantization of open string in background B field and
noncommutative D-brane,''
Nucl.\ Phys.\ B {\bf 568}, 447 (2000)
[hep-th/9906192].


\bibitem{Schwarz:1998qj}
A.~Schwarz,
``Morita equivalence and duality,''
Nucl.\ Phys.\ B {\bf 534}, 720 (1998)
[hep-th/9805034]
\bibitem{Brace:1999ku}
D.~Brace, B.~Morariu and B.~Zumino,
``Dualities of the matrix model from T-duality of the type II string,''
Nucl.\ Phys.\ B {\bf 545}, 192 (1999)
[hep-th/9810099];
``T-duality and Ramond-Ramond backgrounds in the matrix model,''
Nucl.\ Phys.\ B {\bf 549}, 181 (1999)
[hep-th/9811213].


\bibitem{Seiberg:1999vs}
N.~Seiberg and E.~Witten,
``String theory and noncommutative geometry,''
JHEP {\bf 9909}, 032 (1999)
[hep-th/9908142].


\bibitem{DN}
M.~R.~Douglas and N.~A.~Nekrasov,
``Noncommutative field theory,''
Rev.\ Mod.\ Phys.\  {\bf 73}, 977 (2001)
[arXiv:hep-th/0106048].

\bibitem{Sza}
R.~J.~Szabo,
``Quantum field theory on noncommutative spaces,''
Phys.\ Rept.\  {\bf 378}, 207 (2003)
[arXiv:hep-th/0109162].

\bibitem{Bigatti:1999iz}
D.~Bigatti and L.~Susskind,
``Magnetic fields, branes and noncommutative geometry,''
Phys.\ Rev.\ D {\bf 62}, 066004 (2000)
[arXiv:hep-th/9908056].


\bibitem{Yin:1999ba}
Z.~Yin,
``A note on space noncommutativity,''
Phys.\ Lett.\ B {\bf 466}, 234 (1999)
[arXiv:hep-th/9908152].

\bibitem{Kiem:2001pw}
Y.~Kiem, S.~J.~Rey, H.~T.~Sato and J.~T.~Yee,
``Open Wilson lines and generalized star product in nocommutative scalar  field
theories,''
Phys.\ Rev.\ D {\bf 65}, 026002 (2002)
[arXiv:hep-th/0106121].

\bibitem{Kiem:2001fn}
Y.~Kiem, S.~Lee, S.~J.~Rey and H.~T.~Sato,
``Interacting open Wilson lines in noncommutative field theories,''
Phys.\ Rev.\ D {\bf 65}, 046003 (2002)
[arXiv:hep-th/0110215].

\bibitem{Kiem:2001dk}
Y.~Kiem, S.~J.~Rey, H.~T.~Sato and J.~T.~Yee,
``Anatomy of one-loop effective action in noncommutative scalar field
theories,''
Eur.\ Phys.\ J.\ C {\bf 22}, 757 (2002)
[arXiv:hep-th/0107106].

\bibitem{Kiem:2001du}
Y.~j.~Kiem, S.~S.~Kim, S.~J.~Rey and H.~T.~Sato,
``Anatomy of two-loop effective action in noncommutative field theories,''
Nucl.\ Phys.\ B {\bf 641}, 256 (2002)
[arXiv:hep-th/0110066].

\bibitem{Alekseev:1999bs}
A.~Y.~Alekseev, A.~Recknagel and V.~Schomerus,
``Non-commutative world-volume geometries: Branes on SU(2) and fuzzy
spheres,''
JHEP {\bf 9909}, 023 (1999)
[arXiv:hep-th/9908040].


\bibitem{Ishibashi:1988kg}
N.~Ishibashi,
``The Boundary And Crosscap States In Conformal Field Theories,''
Mod.\ Phys.\ Lett.\ A {\bf 4}, 251 (1989).


\bibitem{Cardy:ir}
J.~L.~Cardy,
``Boundary Conditions, Fusion Rules And The Verlinde Formula,''
Nucl.\ Phys.\ B {\bf 324}, 581 (1989).


\bibitem{Klimcik:1996hp}
C.~Klimcik and P.~Severa,
``Open strings and D-branes in WZNW models,''
Nucl.\ Phys.\ B {\bf 488}, 653 (1997)
[arXiv:hep-th/9609112].


\bibitem{Alekseev:1998mc}
A.~Y.~Alekseev and V.~Schomerus,
``D-branes in the WZW model,''
Phys.\ Rev.\ D {\bf 60}, 061901 (1999)
[arXiv:hep-th/9812193].



\bibitem{Alekseev:2000fd}
A.~Y.~Alekseev, A.~Recknagel and V.~Schomerus,
``Brane dynamics in background fluxes and non-commutative geometry,''
JHEP {\bf 0005}, 010 (2000)
[arXiv:hep-th/0003187].


\bibitem{Bachas:2000ik}
C.~Bachas, M.~R.~Douglas and C.~Schweigert,
``Flux stabilization of D-branes,''
JHEP {\bf 0005}, 048 (2000)
[arXiv:hep-th/0003037].

\bibitem{Witten:1983ar}
E.~Witten,
``Nonabelian Bosonization In Two Dimensions,''
Commun.\ Math.\ Phys.\  {\bf 92}, 455 (1984).

\bibitem{McGreevy:2000cw}
J.~McGreevy, L.~Susskind and N.~Toumbas,
``Invasion of the giant gravitons from anti-de Sitter space,''
JHEP {\bf 0006}, 008 (2000)
[arXiv:hep-th/0003075].


\bibitem{Haldane:1983xm}
F.~D.~M.~Haldane,
``Fractional Quantization Of The Hall Effect: A Hierarchy Of Incompressible
Quantum Fluid States,''
Phys.\ Rev.\ Lett.\  {\bf 51}, 605 (1983).



\bibitem{Nair:2001kr}
V.~P.~Nair,
``Gravitational fields on a noncommutative space,''
Nucl.\ Phys.\ B {\bf 651}, 313 (2003)
[arXiv:hep-th/0112114].

\bibitem{Steinacker:2001fj}
H.~Steinacker,
``Quantum field theory on the q-deformed fuzzy sphere,''
arXiv:hep-th/0105126.



\bibitem{Zhang:2001xs}
S.~C.~Zhang and J.~p.~Hu,
``A Four Dimensional Generalization of the Quantum Hall Effect,''
Science {\bf 294}, 823 (2001)
[arXiv:cond-mat/0110572].

\end{thebibliography}
\end{document}